\documentclass[sigconf,natbib=true]{acmart}

\AtBeginDocument{%
  }

\acmConference[SIGIR '25]{}{July 13--18,
 2025}{Padua, Italy}

\usepackage[noend]{algorithmic}
\usepackage[ruled]{algorithm2e}
\usepackage{adjustbox}
\usepackage{enumitem}
\usepackage{multirow}
\usepackage{makecell}
\usepackage{enumitem}
\usepackage{pifont}

\newcommand{\xmark}{\ding{55}}%

\newcommand{\msmarco}{\textsc{MsMarco}\xspace}

\newcommand{\beir}{\textsc{Beir}\xspace}

\newcommand{\seismic}{\textsc{Seismic}\xspace}

\newcommand{\splade}{\textsc{Splade}\xspace}

\newcommand{\spladeT}{\textsc{Splade}-v3\xspace}

\newcommand{\smallmodel}{\textsc{Small}\xspace}
\newcommand{\midmodel}{\textsc{Medium}\xspace}
\newcommand{\bigmodel}{\textsc{Big}\xspace}

\newcommand{\name}{\textsc{Li-Lsr}\xspace}
\newcommand{\nameext}{\emph{Learned Inference-free Retrieval}\xspace}

\copyrightyear{2025}
\acmYear{2025}
\setcopyright{cc}
\setcctype{by}
\acmConference[SIGIR '25]{Proceedings of the 48th International ACM SIGIR Conference on Research and Development in Information Retrieval}{July 13--18, 2025}{Padua, Italy}
\acmBooktitle{Proceedings of the 48th International ACM SIGIR Conference on Research and Development in Information Retrieval (SIGIR '25), July 13--18, 2025, Padua, Italy}
\acmDOI{10.1145/3726302.3730185}
\acmISBN{979-8-4007-1592-1/2025/07}

\begin{document}

\title{Effective Inference-Free Retrieval for Learned Sparse Representations}

\author{Franco Maria Nardini}
\affiliation{%
    \institution{ISTI-CNR}
    \city{Pisa}
    \country{Italy}
}
\orcid{0000-0003-3183-334X}
\email{francomaria.nardini@isti.cnr.it}

\author{Thong Nguyen}

\affiliation{%
    \institution{University of Amsterdam}
    \city{Amsterdam}
    \country{The Netherlands}
}
\orcid{0000-0003-0607-0723}
\email{t.nguyen2@uva.nl}

\author{Cosimo Rulli}
\affiliation{%
    \institution{ISTI-CNR}
    \city{Pisa}
    \country{Italy}
}
\orcid{0000-0003-0194-361X}
\email{cosimo.rulli@isti.cnr.it}

\author{Rossano Venturini}
\affiliation{%
    \institution{University of Pisa}
    \city{Pisa}
    \country{Italy}
}
\orcid{0000-0002-9830-3936}
\email{rossano.venturini@unipi.it}

\author{Andrew Yates}
\affiliation{
\institution{Johns Hopkins University, HLTCOE}
\city{Baltimore}
\country{MD, USA}
}
\orcid{0000-0002-5970-880X}
\email{andrew.yates@jhu.edu}

\renewcommand{\shortauthors}{Nardini \emph{et al.}}

\begin{abstract}
Learned Sparse Retrieval (LSR) is an effective IR approach that exploits pre-trained language models for encoding text into a learned bag of words. Several efforts in the literature have shown that sparsity is key to enabling a good trade-off between the efficiency and effectiveness of the query processor.
To induce the right degree of sparsity, researchers typically use regularization techniques when training LSR models.
Recently, new efficient---inverted index-based---retrieval engines have been proposed, leading to a natural question: has the role of regularization changed in training LSR models? In this paper, we conduct an extended evaluation of regularization approaches for LSR where we discuss their effectiveness, efficiency, and out-of-domain generalization capabilities. 
We first show that regularization can be relaxed to produce more effective LSR encoders. We also show that query encoding is now the bottleneck limiting the overall query processor performance.
To remove this bottleneck, we advance the state-of-the-art of inference-free LSR by proposing \nameext (\name). At training time, \name learns a score for each token, casting the query encoding step into a seamless table lookup. Our approach yields state-of-the-art effectiveness for both in-domain and out-of-domain evaluation, surpassing \spladeT-Doc by $1$ point of mRR@10 on \msmarco and $1.8$ points of nDCG@10 on \beir. 

\end{abstract}

\ccsdesc[500]{Information systems~Retrieval models and ranking}

\keywords{Learned Sparse Retrieval, Inference-Free, Efficiency}


 \maketitle

\vspace{-0.1cm}
\section{Introduction}
\label{sec:intro}

Pretrained Language Models (PLM) have impacted IR in many ways~\cite{DBLP:series/synthesis/2021LinNY}. Indeed, the possibility of producing vector representations, i.e., \emph{embeddings}, of the input text by grasping semantic and contextual information is one of the most important~\cite{karpukhin-etal-2020-dense,xiong2021approximate,khattab2020colbert}. Learning representations allows the problem of retrieving relevant documents for a query to be cast into the problem of finding the closest document embeddings w.r.t. the input query embedding according to a specific similarity metric, which is often dot product or cosine. Learned sparse representations (LSR)~\cite{epic,splade-sigir2021,formal2021splade,formal2022splade,lassance2022efficient-splade, lassance2024splade,zhao2021sparta} map queries and documents into $n$-dimensional vectors where dimensions are grounded on the vocabulary of the collection. When a coordinate is nonzero in a sparse embedding, that indicates that the corresponding term is semantically relevant to the input. 
LSR models have shown to be competitive with dense retrieval models~\cite{DBLP:series/synthesis/2021LinNY,karpukhin-etal-2020-dense,xiong2021approximate,reimers-2019-sentence-bert,khattab2020colbert, santhanam2022colbertv2,daliri2023sampling} and to effectively generalize to out-of-domain datasets~\cite{bruch2023fusion,lassance2024splade}, while being interpretable by design. Moreover, sparse embeddings retain some of the benefits of classical lexical models such as BM25~\cite{bm25original}, including the possibility to use retrieval machinery based on inverted indexes.

Despite the natural link that can be drawn between learned sparse representations and inverted indexes, their matching has only recently been fully realized. In fact, learned sparse representations show important differences w.r.t. normal query/document text, such as longer queries, and the non-Zipfian distribution of terms~\cite{mackenzie2021wacky}, which results in increased per-query latency.
Several efforts have investigated how this performance gap can be closed when using inverted index-based methods. 
In this line, Bruch \emph{et al.} recently introduced \seismic~\cite{bruch2024efficient,bruch2024pairing,bruch2025investigating}, a novel inverted index-based approximate nearest neighbors retrieval, which is shown to outperform state-of-the-art approaches for retrieval with learned sparse representations. \seismic is inherently built to work with Learned Sparse Representations and its properties, such as longer queries, longer documents, and non-Zipfian distribution. 

In this paper, we explore the role of term expansion in Learned Sparse Retrieval, by 
proposing an extended evaluation of regularization approaches for LSR on both in- and out-of-domain benchmarks. We show that regularization can be relaxed in LSR, producing more effective sparse embeddings while still allowing for reduced retrieval times, thanks to modern retrieval engines. In this setting,
 query encoding can be a significant bottleneck on CPU, limiting the performance of the query processor. We thus advance the state of the art by introducing \nameext (\name), a new technique that learns a score for each token, casting the query encoding step to a fast table lookup operation.

In summary, the contributions of this paper are the following:
\begin{itemize}[leftmargin=*]
    \item We present an in-depth evaluation of regularization techniques for LSR, showing its pivotal role in in- and out-of-domain effectiveness. In particular, our setup shows how relaxing regularization improves retrieval quality in both scenarios.
    
    \item We evaluate the consequences of relaxing regularization from an efficiency perspective. The analysis shows that when leveraging modern retrieval engines, even longer (i.e., less sparse) representations can be retrieved efficiently.
    
    \item We propose \nameext (\name), a novel inference-free approach that allows the replacement of the query encoder with a fast lookup table. \name learns the optimal score for each term at training time, enabling both effective and efficient query encoding. \name outperforms state-of-the-art inference-free retrievers such as \spladeT-Doc, with a margin of $1$ point of mRR on \msmarco and $1.8$ points of nDCG@10 on \beir. 
\end{itemize}



\section{Methodology}
\label{sec:methodology}
In this section, we define the methodology used to assess the impact of term expansion in modern LSR. Moreover, we introduce \name, a novel approach to the inference-free paradigm that learns the importance score of each term at training time.

\subsection{Exploring Term Expansion in LSR}
\label{subsec:expexp}
Unlike na\"ive term-matching engines, which exploit terms that appear in the input text, learned sparse representations are trained to activate relevant terms regardless of whether they appear in the input tokens. In fact, LSR can encompass \emph{term expansion} on both the query and the document, i.e., adding terms to the representation to improve its generalization power. Term expansion in LSR encoders is enabled and controlled by regularization terms applied to the loss function, which control the sparsity in the representations, thus enabling a practical---yet effective---use of the obtained embeddings. 

In this section, we define a methodology that aims to assess the effectiveness of LSR as a function of the regularization applied during training. Equation \ref{dist} introduces a general loss function $\mathcal{L}$ used for learning sparse representations. 
\begin{equation}
\label{dist}
\mathcal{L} \ = \ \mathcal{L}_{\text{rank}} + \lambda_q \, \mathcal{L}^{Q}_{\text{reg}} \ + \  \lambda_d \,\mathcal{L}^{D}_{\text{reg}}
\end{equation}

In state-of-the-art methods, $\mathcal{L}_{\text{rank}}$ is typically a distillation loss~\cite{formal2022splade,lassance2024splade,lassance2022efficient-splade}, encouraging the student model to align its scores with a strong teacher like a cross-encoder~\cite{wang2020minilm}.

We investigate how $\lambda_q$ and $\lambda_d$ impact the effectiveness of an LSR model and how they interact with different loss functions. To do so, we opt for the following methodology. We first define a training configuration as a pair comprising a training loss and a regularizer. We then build three LSR models: \smallmodel, \midmodel, and \bigmodel, which are categorized based on the average number of non-zero values in the document representations for the collection, which we define as $|D|$. In \smallmodel models, we do not employ term expansion, i.e., $|D|$ is comparable to the number of tokens per document. For \msmarco, this number is approximately $60$. In \midmodel models, we set term expansion to double the number of non-zero values in the documents compared to the \smallmodel configuration. This aligns 
with the expansion achieved by state-of-the-art models such as \splade (119) and \spladeT (168)~\cite{lassance2024splade}. For \bigmodel models, we set $|D|$ to be roughly $5\times$ larger than the non-expanded scenario. This setting is particularly important. Expansion on both the query and document sides has traditionally been constrained to ensure efficient retrieval using classical inverted indexes. Our evaluation assesses how the retrieval effectiveness benefits from a larger expansion. Moreover, \bigmodel aligns with most common dense representations~\cite{lin2023train} regarding embedding memory footprint. A non-zero entry in a sparse embedding requires $16$ bits for the token identifier, as long as the vocabulary size is smaller than $2^{16}$, plus $16$ bits for the token weight when using float16 precision. When storing dense vectors using a float16 per dimension, a sparse entry requires twice the memory of a dense one. Given that most common BERT-based dense encoders produce $768$-dimensional vectors \cite{lin2023train,hofstatter2021efficiently}, the memory footprint of embeddings generated by \bigmodel is approximately the same as those produced by a dense encoder, i.e., $1536$ bytes per vector.

\subsection{Learned Inference-Free Retrieval}
\label{subsec:lif}
The \emph{inference-free} paradigm bypasses the costly query encoding phase by assigning a fixed value, such as $1$, to each query term. While eliminating encoding costs brings efficiency gains, it significantly reduces retrieval effectiveness. For example, as noted by Geng \emph{et al.}~\cite{geng2024towards}, \spladeT experiences a drop of $2.4$ points in mRR@10 on \msmarco and $4.7$ points in nDCG@10 on \beir when query encoding is removed~\cite{geng2024towards}.

We propose to advance the state of the art in inference-free retrieval by introducing \nameext (\name). \name learns a static relevance score for each token during training by projecting the output of word embeddings into a scalar value using a simple linear layer. This transforms query encoding into a seamless lookup in a term-score dictionary. Unlike existing methods that rely on fixed scores or lexical statistics, \name enables the model to learn the optimal value for each token at training time.

Let us consider a tokenizer that maps a query input sequence $q$ into a set of tokens $x = \{x_0, \dots, x_l \}$, where $l$ is the number of tokens and $x_i \in \mathrm{I}_{V}$, being $\mathrm{I}_{V}$ the set of valid indices in $V$. Our goal is to learn a mapping  
$
S = \{ x_i : s_i\}_{i=0}^{|V|}
$
to reduce query encoding to a simple lookup in $S$.  

Transformer-based encoders rely on an embedding module, $E$, to project $x$ into a vector representation $E(x) \in \mathbb{R}^{l \times d}$, where $d$ is the vector dimensionality. Typically, $E$ consists of multiple components, including but not limited to word embeddings, positional embeddings, and a normalization layer. In our approach, we consider only the word embedding module, $E_W$, because it provides a \emph{context-independent} representation for a given token $x_i$, namely  
$
E_W(x) = [ E_W(x_1), \dots, E_W(x_l) ].
$  
In contrast, positional embeddings depend on token position,  $E_P(x_i) = E_P(x_i; i)$, and layer normalization requires the computation over the entire input sequence.

We introduce a linear layer $w$ with a bias $b$ to project the output of the word embedding module into a per-token score, and then we apply a positivity constraint, similarly to the \splade log-saturation function, to obtain $s_i$:
\begin{equation}
   s_i = \log (1 + \text{ReLU} [w^T   E_W (x_i) + b]).
\end{equation}
Whether the same token appears more than once in the input sequence, the final score of the token is multiplied by the number of times it appears.

\begin{table}[!tb]
    \centering
    \adjustbox{max width=\columnwidth}{
    \begin{tabular}{lllrrrrrr}
        \toprule
         Loss & Reg. & Size  &$\lambda_q$ & $\lambda_d$ & |Q| & |D|& \msmarco & \beir \\
         \midrule
         \multirow{6}{*}{$KL$} & \multirow{3}{*}{$L_1$} & 
         \smallmodel & 1e-3 & 1e-4& 8 & 73 & 39.08 & 48.83  \\
          & & \midmodel & 5e-5 & 5e-5 & 14 & 128 & 39.44 & 49.55  \\
          & & \bigmodel & 1e-5 &  1e-5 & 16 & 304 &\textbf{39.81}& 49.04 \\
        \cmidrule{2-9}
         & \multirow{3}{*}{FLOPs} & 
           \smallmodel & 1e-2 & 1e-2 &  17 & 78 & 38.49  & 48.52\\
          & & \midmodel & 2e-3& 2e-3  &14 &  145&38.91 & 50.02 \\
          & & \bigmodel & 2e-4 & 2e-4 & 14 & 355 & 39.51 &  49.24 \\
        \midrule
       \multirow{6}{*}{MSE}   
        & \multirow{3}{*}{$L_1$} & \smallmodel & 5e-2 & 5e-2  & 13 & 38 & 37.83  & 48.07 \\
        & & \midmodel & 5e-3 & 5e-3 & 30 & 133 &  38.68 & 50.71\\
        & & \bigmodel & 1e-3 & 1e-3 & 45 & 326 &  38.86 & 50.65\\
        \cmidrule{2-9}
        & \multirow{3}{*}{FLOPs} & \smallmodel & 2.50 & 2.50 & 34 & 69 & 37.64 & 47.30 \\
       & & \midmodel  & 5e-1& 5e-1 & 43 & 138 & 38.38 & 50.35 \\
       & & \bigmodel & 5e-2 & 5e-2 & 38 &  294& 38.84 & \textbf{51.02} \\
         \bottomrule
    \end{tabular}}
    \caption{Performance of an LSR Model with different levels of expansion. Best results in terms of mRR@10 and nDCG@10 are reported in bold.}
    \label{tab:main_table}
\end{table}

\section{Experimental Evaluation}
\label{sec:exps}


\smallskip
\noindent \textbf{Datasets and Evaluation}. We train our models, i.e., \smallmodel, \midmodel, \bigmodel, on the \msmarco~\cite{nguyen2016msmarco} dataset over four different combinations of training loss, i.e., Kullback-Leibler (KL) or Mean Squared Error (MSE), and regularization strategies ($L_1$ or FLOPs), distilling from the \emph{msmarco-hard-negatives}.\footnote{\url{https://huggingface.co/datasets/sentence-transformers/msmarco-hard-negatives}} We use a single negative for each query. We employ a batch size of $128$, a learning rate of 5e-5, and we train for $150$,$000$ steps. We start our training from the Co-Condenser~\cite{gao2022unsupervised} checkpoint available on HuggingFace.\footnote{\url{https://huggingface.co/Luyu/co-condenser-marco}} We evaluate our LSR models both on in-domain and out-of-domain benchmarks. For the former, we use the $6$,$980$ queries in the \emph{dev/small} set of \msmarco, evaluating the ranking performance using mRR@10. For the latter, we use the $13$ datasets in the \beir~\cite{thakur2021beir} collections, and we adopt nDCG@10 as a metric, following~\cite{lassance2024splade, geng2024towards}.

\vspace{1mm}
\noindent \textbf{Indexing and Retrieval}. We index our collection using \seismic~\cite{bruch2024efficient}. \seismic allows both for exhaustive and approximate search. We employ exhaustive search in effectiveness-oriented evaluations. We resort to approximate search when comparing the efficiency-effectiveness trade-off achieved by different LSR models. We perform a grid search over the length of the truncated posting lists of \seismic, varying it in $\{2000, 4000, 6000, 8000\}$. Following Bruch \emph{et al.} \cite{bruch2025investigating}, we set the summary energy, i.e., $\alpha$, to $0.4$ and the number of centroids per list to one-tenth of the posting list length. The efficiency evaluation was conducted on a machine equipped with two Intel Xeon Silver 4314 CPU, clocked at 2.40GHz, 256GiB of RAM. All the experiments involving retrieval and encoding were conducted using a single execution thread.

\subsection{Impact of Term Expansion in LSR}
\label{subsec:exp1}
We present the results of our comparison in Table~\ref{tab:main_table}. 
For each row, we report the type of loss function used for training (KL or MSE), the regularization method ($L_1$ or FLOPs), and the corresponding values of query and document regularization ($\lambda_q$ and $\lambda_d$). Additionally, we provide the resulting number of non-zero entries per query ($|Q|$) and per document ($|D|$), along with their retrieval performance on \msmarco (mRR@10) and \beir (nDCG@10).

Table~\ref{tab:main_table} shows that both query and document density positively correlate with effectiveness, particularly for documents. In fact, increasing document length proves to be an effective strategy for improving the retrieval quality of the encoder.  
The only exception is nDCG@10 on \beir for models trained with $KL$, where performance peaks with \midmodel models rather than \bigmodel ones. We also observe that $|Q|$ is influenced not only by $\lambda_q$ but also by $\lambda_d$. This indicates that even when query regularization is disabled ($\lambda_q = 0$), query expansion remains constrained. Additionally, we find that models perform best when $\lambda_q$ and $\lambda_d$ are equal, with the exception of \smallmodel in the $L_1 + KL$ scenario. 

Regarding in-domain and out-of-domain effectiveness, the $KL$ divergence achieves the best results on \msmarco, reaching a maximum mRR@10 of $39.81$ with $L_1$ regularization at the maximum allowed expansion.  
The \smallmodel model, trained using the same configuration, attains an mRR@10 of $39.08$, while offering the advantage of requiring $2\times$ and $4\times$ smaller embedding collections compared to \midmodel and \bigmodel, respectively.  
On the other hand, MSE is more effective for out-of-domain retrieval, particularly when combined with FLOPs regularization. The best-performing model is obtained using FLOPs regularization and MSE as the training loss, achieving an nDCG@10 of $51.02$ on \beir. We attribute this to the larger query expansion induced by MSE loss compared to $KL$ divergence.

\subsection{Efficiency-Effectiveness Trade-off}
\begin{figure}[b]
    \centering
    \includegraphics[width=1.\linewidth]{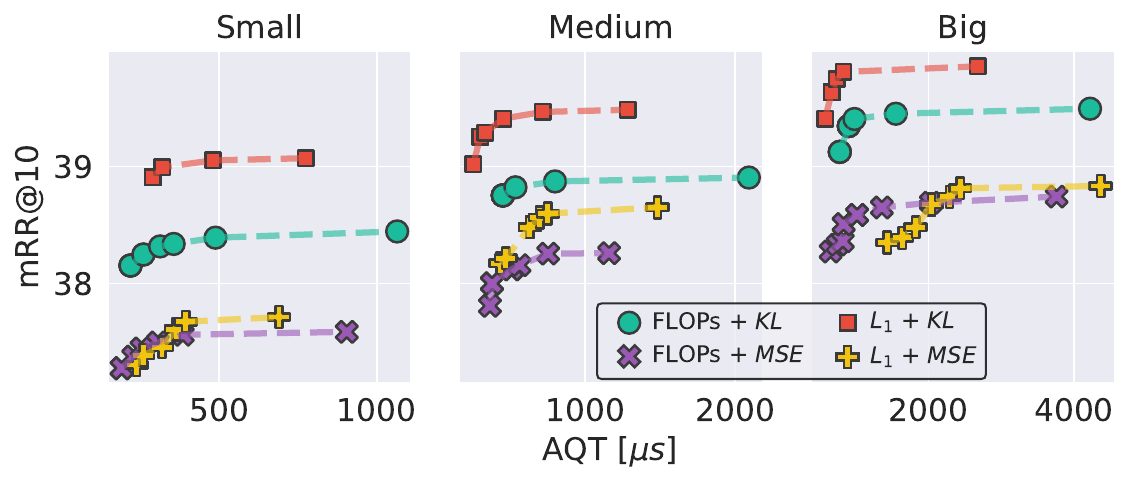}
    \caption{Efficiency-effectiveness trade-off (AQT vs. mRR@10) of different LSR models with different expansion.}
    \label{fig:mrr_aqt}
\end{figure}

In this section, we investigate the relationship between expansion and retrieval efficiency. Figure~\ref{fig:mrr_aqt} reports the mRR@10 on \msmarco as a function of the average query time (AQT, in $\mu$sec.) for each of the models in Table~\ref{tab:main_table}, grouped into \smallmodel, \midmodel, and \bigmodel.

Despite the large number of non-zero terms in queries and documents, models from the \bigmodel family still enable highly efficient retrieval. In particular, the model trained with $KL$ divergence as the loss function and $L_1$ regularization achieves an mRR@10 of $39.80$—over $99.9\%$ of the peak effectiveness—in approximately $800$ $\mu$sec. using single-threaded execution.  
Notably, this level of mRR@10 is comparable to that of multi-vector encoders such as ColBERTv2~\cite{santhanam2022colbertv2}.
As an example, we tested the optimized EMVB~\cite{nardini2024efficient} retrieval engine in the $k=10$, $m=32$ setting; while yielding a mRR@10 of $39.7$, retrieving with EMVB takes about $60$ $m$sec., almost two orders of magnitude slower than \seismic.


As mentioned above, query expansion plays a key role in out-of-domain effectiveness. While a decrease in query sparsity hinders the usage of standard inverted indexes~\cite{bruch2024efficient}, \seismic's efficiency is resilient to query expansion.
For example, the FLOPs + MSE model can quickly reach $99.5\%$ of its maximum performance in $1.3$ $m$sec. 

\smallskip
\noindent\textbf{Comparison with encoding time}
We complement the efficiency-effectiveness trade-off analysis by measuring the time required to encode a query with a BERT-based model~\cite{kenton2019bert}. We measure the inference time using the \texttt{rust-bert} Rust package~\cite{becquin2020end}. To align the evaluation with the methodology used for retrieval, we measure it on the CPU in single-thread execution with a batch size of $1$. We employ \msmarco \emph{dev/small} as a benchmark. The average encoding time is $24$ $m$sec., one order of magnitude slower than almost-exact retrieval with \seismic. This evidence poses the interesting challenge of reducing the query encoding time for sparse retrievers. 

\begin{table}[]
    \centering
    \adjustbox{max width=\columnwidth}{
    \begin{tabular}{llllrrrr}
        \toprule
        Model & Neg. & Teach. & PT & \textsc{MM} & \textsc{DL19}&\textsc{DL20} & \beir \\ 
        \midrule
        \splade-Doc-Distill & 1 &1 & \xmark &  36.5 & 69.8  & -&45.0 \\
        \spladeT-Doc& M & M & \xmark & 37.8 & 71.5 & 70.3 & 47.0 \\
        Geng \emph{et al}.~\cite{geng2024towards} &M & M & $\checkmark$  & 37.8 & 72.1& - & 50.4 \\
        \midrule 
        \textbf{\name} \\
        \cmidrule{1-1}
        \midmodel & 1 & 1 & \xmark &  37.6 & 69.6  & 69.6 &47.6\\
          \bigmodel & 1 & 1 & \xmark & 38.8 & 72.1  & 70.5 &48.8 \\
        \midrule
        \textbf{\name + IDF} \\ 
        \cmidrule{1-1}\midmodel & 1 & 1 & \xmark &  37.8 & 71.3  & 68.3& 48.2\\
           \bigmodel & 1 & 1 & \xmark & 38.0 & 71.6 & 69.6& 48.9 \\
        \bottomrule
    \end{tabular}}
    \caption{Comparison between inference-free LSR models.\vspace{-5mm}}
    \label{tab:lilsr}
\end{table}

\vspace{-0.1cm}
\subsection{Learned Inference-Free Retrieval}
\label{subsec:explilsr}
We now evaluate the effectiveness of our \emph{Learned Inference-Free Retrieval} (\name) technique, which aims at introducing a learned inference-free query encoding. We compare our approach with state-of-the-art inference-free retrievers in Table~\ref{tab:lilsr}, including \splade-Doc-Distill~\cite{formal2022splade}, \spladeT-Doc~\cite{lassance2024splade}, and a method by Geng \emph{et al.}~\cite{geng2024towards} that uses inverse document frequency as query term weights. For each model, we report the in-domain performance measured as mRR@10 on \msmarco (``\textsc{MM}''), and as nDCG@10 on \textsc{TrecDL}-2019 (``DL19'') and \textsc{TrecDL}-2020 (``DL20''). The out-of-domain effectiveness is measured in terms of nDCG@10 on \beir. We also describe its training methodology by specifying the number of negatives used per query (``Neg.'') and whether it employs single or multiple teachers (``Teach.''). To this regard, we highlight that Geng \emph{et al.}~\cite{geng2024towards} use a vast pre-training on different data sources before fine-tuning on \msmarco~\cite{geng2024towards}. Our approach is trained with the settings detailed in Section~\ref{sec:exps}, namely with a single teacher and a single negative per query. We use $L_1$ as the regularization function and $KL$ as loss. In our experiments, we observed that the $KL$ loss yields consistently better results than MSE for \name. Capitalizing on the lessons learned in Section~\ref{subsec:exp1}, we train models with different document expansions. We report the results of \midmodel ($\lambda_d$ = $1$e$-4$) and \bigmodel ($\lambda_d$ = $1$e$-5$), the former to assess the effectiveness of \name compared to other inference-free techniques, the latter to explore the potential of inference-free retrieval when paired with boosted document expansion.

Table~\ref{tab:lilsr} shows the effectiveness of \name. Our \name - \midmodel model rivals the performance of \spladeT-Doc for in-domain and surpasses it for out-of-domain, despite \spladeT-Doc employing multiple teachers and multiple negatives per query. With the same training budget, \name-\midmodel outperforms \splade-Doc-Distill by $1.1$ points of mRR@10 on \msmarco and $2.6$ of nDCG@10 on \beir. \name - \bigmodel surpasses by $1.0$ points both \spladeT and the approach by Geng \emph{et al.}~\cite{geng2024towards}, despite the vast disparity between the training budget. This showcases the effectiveness of our approach to learning the scores to assign to query terms rather than relying on fixed, predetermined values. We also combine our approach with the one from Geng \emph{et al.}~\cite{geng2024towards} by computing the query term scores by multiplying our learned word embedding projection with the IDF value of the corresponding token. These configurations are reported as ``\name + IDF''. Our experiments show that combining IDF with our learned approach allows smaller models (\name + IDF - \midmodel) to improve the out-of-domain-effectiveness. In contrast, the large model, i.e., \name + IDF - \bigmodel, poorly benefits from it.


\section{Conclusions and Future Work}
\label{sec:concl}

We presented a study about the role of term expansion in sparse encoders, both in terms of effectiveness and efficiency. We found that a lighter regularization---allowing more non-zero terms in queries and documents---enables for improved retrieval quality. Thanks to new state-of-the-art retrieval engines explicitly tailored for LSR, even with light regularization, retrieval can be done in a few milliseconds per query. These considerations facilitate more effective sparse encoders that rival multi-vector representations in effectiveness while being almost two orders of magnitude faster. Moreover, we introduce \name, a new method for inference-free retrieval. \name enables the encoding phase to be converted into a fast dictionary lookup of a learned score for each token. Our approach outperforms existing inference-free methods, even when relying on much simpler training recipes. For example, it outperforms \spladeT-Doc by $1$ point of mRR on \msmarco and $1.8$ points of nDCG@10 on \beir.
As future work, given the importance of document expansion highlighted by this analysis, we intend to investigate compression techniques for learned sparse representation.

\smallskip
\noindent \textbf{Acknowledgments.}
We acknowledge the support of ``FAIR - Future Artificial Intelligence Research'' - Spoke 1 ''Human-centered AI'' funded by the European Union (EU) under the NextGeneration EU programme, and by the Horizon Europe RIA ``Extreme Food Risk Analytics'' (EFRA) funded by the European Commission under the NextGeneration EU programme grant agreement n. 101093026. Views and opinions expressed are however those of the authors only and do not necessarily reflect those of the EU or European Commission-EU. Neither the EU nor the granting authority can be held responsible for them.
SoBigData.it receives funding from European Union - NextGenerationEU - National Recovery and Resilience Plan (Piano Nazionale di Ripresa e Resilienza, PNRR) - Project: “SoBigData.it - Strengthening the Italian RI for Social Mining and Big Data Analytics” - Prot. IR0000013 - Avviso n. 3264 del S28/12/2021.
This research was partially supported by the Dutch Research Council (NWO) under project number VI.Vidi.223.166.



\bibliographystyle{ACM-Reference-Format}
\bibliography{sample-base}


\begin{thebibliography}{30}


\ifx \showCODEN    \undefined \def \showCODEN     #1{\unskip}     \fi
\ifx \showISBNx    \undefined \def \showISBNx     #1{\unskip}     \fi
\ifx \showISBNxiii \undefined \def \showISBNxiii  #1{\unskip}     \fi
\ifx \showISSN     \undefined \def \showISSN      #1{\unskip}     \fi
\ifx \showLCCN     \undefined \def \showLCCN      #1{\unskip}     \fi
\ifx \shownote     \undefined \def \shownote      #1{#1}          \fi
\ifx \showarticletitle \undefined \def \showarticletitle #1{#1}   \fi
\ifx \showURL      \undefined \def \showURL       {\relax}        \fi
\providecommand\bibfield[2]{#2}
\providecommand\bibinfo[2]{#2}
\providecommand\natexlab[1]{#1}
\providecommand\showeprint[2][]{arXiv:#2}

\bibitem[Becquin(2020)]%
        {becquin2020end}
\bibfield{author}{\bibinfo{person}{Guillaume Becquin}.}
  \bibinfo{year}{2020}\natexlab{}.
\newblock \showarticletitle{End-to-end {NLP} Pipelines in Rust}. In
  \bibinfo{booktitle}{\emph{Proceedings of Second Workshop for NLP Open Source
  Software (NLP-OSS)}}. \bibinfo{publisher}{Association for Computational
  Linguistics}, \bibinfo{pages}{20--25}.
\newblock
\urldef\tempurl%
\url{https://www.aclweb.org/anthology/2020.nlposs-1.4}
\showURL{%
\tempurl}


\bibitem[Bruch et~al\mbox{.}(2023)]%
        {bruch2023fusion}
\bibfield{author}{\bibinfo{person}{Sebastian Bruch}, \bibinfo{person}{Siyu
  Gai}, {and} \bibinfo{person}{Amir Ingber}.} \bibinfo{year}{2023}\natexlab{}.
\newblock \showarticletitle{An Analysis of Fusion Functions for Hybrid
  Retrieval}.
\newblock \bibinfo{journal}{\emph{ACM Transactions on Information Systems}}
  \bibinfo{volume}{42}, \bibinfo{number}{1}, Article \bibinfo{articleno}{20}
  (\bibinfo{date}{August} \bibinfo{year}{2023}), \bibinfo{numpages}{35}~pages.
\newblock


\bibitem[Bruch et~al\mbox{.}(2024a)]%
        {bruch2024efficient}
\bibfield{author}{\bibinfo{person}{Sebastian Bruch},
  \bibinfo{person}{Franco~Maria Nardini}, \bibinfo{person}{Cosimo Rulli}, {and}
  \bibinfo{person}{Rossano Venturini}.} \bibinfo{year}{2024}\natexlab{a}.
\newblock \showarticletitle{Efficient inverted indexes for approximate
  retrieval over learned sparse representations}. In
  \bibinfo{booktitle}{\emph{Proceedings of the 47th International ACM SIGIR
  Conference on Research and Development in Information Retrieval}}.
  \bibinfo{pages}{152--162}.
\newblock


\bibitem[Bruch et~al\mbox{.}(2024b)]%
        {bruch2024pairing}
\bibfield{author}{\bibinfo{person}{Sebastian Bruch},
  \bibinfo{person}{Franco~Maria Nardini}, \bibinfo{person}{Cosimo Rulli}, {and}
  \bibinfo{person}{Rossano Venturini}.} \bibinfo{year}{2024}\natexlab{b}.
\newblock \showarticletitle{Pairing Clustered Inverted Indexes with $\kappa$-NN
  Graphs for Fast Approximate Retrieval over Learned Sparse Representations}.
  In \bibinfo{booktitle}{\emph{Proceedings of the 33rd ACM International
  Conference on Information and Knowledge Management}}.
  \bibinfo{pages}{3642--3646}.
\newblock


\bibitem[Bruch et~al\mbox{.}(2025)]%
        {bruch2025investigating}
\bibfield{author}{\bibinfo{person}{Sebastian Bruch},
  \bibinfo{person}{Franco~Maria Nardini}, \bibinfo{person}{Cosimo Rulli},
  \bibinfo{person}{Rossano Venturini}, {and} \bibinfo{person}{Leonardo
  Venuta}.} \bibinfo{year}{2025}\natexlab{}.
\newblock \showarticletitle{Investigating the Scalability of Approximate Sparse
  Retrieval Algorithms to Massive Datasets}. In
  \bibinfo{booktitle}{\emph{European Conference on Information Retrieval}}.
  Springer, \bibinfo{pages}{437--445}.
\newblock


\bibitem[Daliri et~al\mbox{.}(2023)]%
        {daliri2023sampling}
\bibfield{author}{\bibinfo{person}{Majid Daliri}, \bibinfo{person}{Juliana
  Freire}, \bibinfo{person}{Christopher Musco}, \bibinfo{person}{Aécio
  Santos}, {and} \bibinfo{person}{Haoxiang Zhang}.}
  \bibinfo{year}{2023}\natexlab{}.
\newblock \bibinfo{title}{Sampling Methods for Inner Product Sketching}.
\newblock
\showeprint[arxiv]{2309.16157}~[cs.DB]


\bibitem[Formal et~al\mbox{.}(2021a)]%
        {formal2021splade}
\bibfield{author}{\bibinfo{person}{Thibault Formal}, \bibinfo{person}{Carlos
  Lassance}, \bibinfo{person}{Benjamin Piwowarski}, {and}
  \bibinfo{person}{Stéphane Clinchant}.} \bibinfo{year}{2021}\natexlab{a}.
\newblock \bibinfo{title}{SPLADE v2: Sparse Lexical and Expansion Model for
  Information Retrieval}.
\newblock
\showeprint[arxiv]{2109.10086}~[cs.IR]


\bibitem[Formal et~al\mbox{.}(2022)]%
        {formal2022splade}
\bibfield{author}{\bibinfo{person}{Thibault Formal}, \bibinfo{person}{Carlos
  Lassance}, \bibinfo{person}{Benjamin Piwowarski}, {and}
  \bibinfo{person}{St\'{e}phane Clinchant}.} \bibinfo{year}{2022}\natexlab{}.
\newblock \showarticletitle{From Distillation to Hard Negative Sampling: Making
  Sparse Neural IR Models More Effective}. In
  \bibinfo{booktitle}{\emph{Proceedings of the 45th International ACM SIGIR
  Conference on Research and Development in Information Retrieval}} (Madrid,
  Spain). \bibinfo{pages}{2353--2359}.
\newblock


\bibitem[Formal et~al\mbox{.}(2021b)]%
        {splade-sigir2021}
\bibfield{author}{\bibinfo{person}{Thibault Formal}, \bibinfo{person}{Benjamin
  Piwowarski}, {and} \bibinfo{person}{St\'{e}phane Clinchant}.}
  \bibinfo{year}{2021}\natexlab{b}.
\newblock \showarticletitle{SPLADE: Sparse Lexical and Expansion Model for
  First Stage Ranking}. In \bibinfo{booktitle}{\emph{Proceedings of the 44th
  International ACM SIGIR Conference on Research and Development in Information
  Retrieval}} (Virtual Event, Canada). \bibinfo{pages}{2288--2292}.
\newblock


\bibitem[Gao and Callan(2022)]%
        {gao2022unsupervised}
\bibfield{author}{\bibinfo{person}{Luyu Gao} {and} \bibinfo{person}{Jamie
  Callan}.} \bibinfo{year}{2022}\natexlab{}.
\newblock \showarticletitle{Unsupervised Corpus Aware Language Model
  Pre-training for Dense Passage Retrieval}. In
  \bibinfo{booktitle}{\emph{Proceedings of the 60th Annual Meeting of the
  Association for Computational Linguistics (Volume 1: Long Papers)}}.
  \bibinfo{pages}{2843--2853}.
\newblock


\bibitem[Geng et~al\mbox{.}(2024)]%
        {geng2024towards}
\bibfield{author}{\bibinfo{person}{Zhichao Geng}, \bibinfo{person}{Dongyu Ru},
  {and} \bibinfo{person}{Yang Yang}.} \bibinfo{year}{2024}\natexlab{}.
\newblock \showarticletitle{Towards Competitive Search Relevance For
  Inference-Free Learned Sparse Retrievers}.
\newblock \bibinfo{journal}{\emph{arXiv preprint arXiv:2411.04403}}
  (\bibinfo{year}{2024}).
\newblock


\bibitem[Hofst{\"a}tter et~al\mbox{.}(2021)]%
        {hofstatter2021efficiently}
\bibfield{author}{\bibinfo{person}{Sebastian Hofst{\"a}tter},
  \bibinfo{person}{Sheng-Chieh Lin}, \bibinfo{person}{Jheng-Hong Yang},
  \bibinfo{person}{Jimmy Lin}, {and} \bibinfo{person}{Allan Hanbury}.}
  \bibinfo{year}{2021}\natexlab{}.
\newblock \showarticletitle{Efficiently teaching an effective dense retriever
  with balanced topic aware sampling}. In \bibinfo{booktitle}{\emph{Proceedings
  of the 44th International ACM SIGIR Conference on Research and Development in
  Information Retrieval}}. \bibinfo{pages}{113--122}.
\newblock


\bibitem[Karpukhin et~al\mbox{.}(2020)]%
        {karpukhin-etal-2020-dense}
\bibfield{author}{\bibinfo{person}{Vladimir Karpukhin}, \bibinfo{person}{Barlas
  Oguz}, \bibinfo{person}{Sewon Min}, \bibinfo{person}{Patrick Lewis},
  \bibinfo{person}{Ledell Wu}, \bibinfo{person}{Sergey Edunov},
  \bibinfo{person}{Danqi Chen}, {and} \bibinfo{person}{Wen-tau Yih}.}
  \bibinfo{year}{2020}\natexlab{}.
\newblock \showarticletitle{Dense Passage Retrieval for Open-Domain Question
  Answering}. In \bibinfo{booktitle}{\emph{Proceedings of the 2020 Conference
  on Empirical Methods in Natural Language Processing}}.
  \bibinfo{pages}{6769--6781}.
\newblock


\bibitem[Kenton and Toutanova(2019)]%
        {kenton2019bert}
\bibfield{author}{\bibinfo{person}{Jacob Devlin Ming-Wei~Chang Kenton} {and}
  \bibinfo{person}{Lee~Kristina Toutanova}.} \bibinfo{year}{2019}\natexlab{}.
\newblock \showarticletitle{Bert: Pre-training of deep bidirectional
  transformers for language understanding}. In
  \bibinfo{booktitle}{\emph{Proceedings of naacL-HLT}},
  Vol.~\bibinfo{volume}{1}. Minneapolis, Minnesota.
\newblock


\bibitem[Khattab and Zaharia(2020)]%
        {khattab2020colbert}
\bibfield{author}{\bibinfo{person}{Omar Khattab} {and} \bibinfo{person}{Matei
  Zaharia}.} \bibinfo{year}{2020}\natexlab{}.
\newblock \showarticletitle{Colbert: Efficient and effective passage search via
  contextualized late interaction over bert}. In
  \bibinfo{booktitle}{\emph{Proceedings of the 43rd International ACM SIGIR
  conference on research and development in Information Retrieval}}.
  \bibinfo{pages}{39--48}.
\newblock


\bibitem[Lassance and Clinchant(2022)]%
        {lassance2022efficient-splade}
\bibfield{author}{\bibinfo{person}{Carlos Lassance} {and}
  \bibinfo{person}{St\'{e}phane Clinchant}.} \bibinfo{year}{2022}\natexlab{}.
\newblock \showarticletitle{An Efficiency Study for SPLADE Models}. In
  \bibinfo{booktitle}{\emph{Proceedings of the 45th International ACM SIGIR
  Conference on Research and Development in Information Retrieval}} (Madrid,
  Spain). \bibinfo{pages}{2220--2226}.
\newblock


\bibitem[Lassance et~al\mbox{.}(2024)]%
        {lassance2024splade}
\bibfield{author}{\bibinfo{person}{Carlos Lassance}, \bibinfo{person}{Herv{\'e}
  D{\'e}jean}, \bibinfo{person}{Thibault Formal}, {and}
  \bibinfo{person}{St{\'e}phane Clinchant}.} \bibinfo{year}{2024}\natexlab{}.
\newblock \showarticletitle{SPLADE-v3: New baselines for SPLADE}.
\newblock \bibinfo{journal}{\emph{arXiv preprint arXiv:2403.06789}}
  (\bibinfo{year}{2024}).
\newblock


\bibitem[Lin et~al\mbox{.}(2021)]%
        {DBLP:series/synthesis/2021LinNY}
\bibfield{author}{\bibinfo{person}{Jimmy Lin},
  \bibinfo{person}{Rodrigo~Frassetto Nogueira}, {and} \bibinfo{person}{Andrew
  Yates}.} \bibinfo{year}{2021}\natexlab{}.
\newblock \bibinfo{booktitle}{\emph{Pretrained Transformers for Text Ranking:
  {BERT} and Beyond}}.
\newblock \bibinfo{publisher}{Morgan {\&} Claypool Publishers}.
\newblock


\bibitem[Lin et~al\mbox{.}(2023)]%
        {lin2023train}
\bibfield{author}{\bibinfo{person}{Sheng-Chieh Lin}, \bibinfo{person}{Akari
  Asai}, \bibinfo{person}{Minghan Li}, \bibinfo{person}{Barlas Oguz},
  \bibinfo{person}{Jimmy Lin}, \bibinfo{person}{Yashar Mehdad},
  \bibinfo{person}{Wen-tau Yih}, {and} \bibinfo{person}{Xilun Chen}.}
  \bibinfo{year}{2023}\natexlab{}.
\newblock \showarticletitle{How to Train Your Dragon: Diverse Augmentation
  Towards Generalizable Dense Retrieval}. In \bibinfo{booktitle}{\emph{Findings
  of the Association for Computational Linguistics: EMNLP 2023}}.
  \bibinfo{pages}{6385--6400}.
\newblock


\bibitem[MacAvaney et~al\mbox{.}(2020)]%
        {epic}
\bibfield{author}{\bibinfo{person}{Sean MacAvaney},
  \bibinfo{person}{Franco~Maria Nardini}, \bibinfo{person}{Raffaele Perego},
  \bibinfo{person}{Nicola Tonellotto}, \bibinfo{person}{Nazli Goharian}, {and}
  \bibinfo{person}{Ophir Frieder}.} \bibinfo{year}{2020}\natexlab{}.
\newblock \showarticletitle{Expansion via Prediction of Importance with
  Contextualization}. In \bibinfo{booktitle}{\emph{Proceedings of the 43rd
  International ACM SIGIR Conference on Research and Development in Information
  Retrieval}} (Virtual Event, China). \bibinfo{pages}{1573--1576}.
\newblock


\bibitem[Mackenzie et~al\mbox{.}(2021)]%
        {mackenzie2021wacky}
\bibfield{author}{\bibinfo{person}{Joel Mackenzie}, \bibinfo{person}{Andrew
  Trotman}, {and} \bibinfo{person}{Jimmy Lin}.}
  \bibinfo{year}{2021}\natexlab{}.
\newblock \bibinfo{title}{Wacky Weights in Learned Sparse Representations and
  the Revenge of Score-at-a-Time Query Evaluation}.
\newblock
\showeprint[arxiv]{2110.11540}~[cs.IR]


\bibitem[Nardini et~al\mbox{.}(2024)]%
        {nardini2024efficient}
\bibfield{author}{\bibinfo{person}{Franco~Maria Nardini},
  \bibinfo{person}{Cosimo Rulli}, {and} \bibinfo{person}{Rossano Venturini}.}
  \bibinfo{year}{2024}\natexlab{}.
\newblock \showarticletitle{Efficient Multi-vector Dense Retrieval with Bit
  Vectors}. In \bibinfo{booktitle}{\emph{European Conference on Information
  Retrieval}}. Springer, \bibinfo{pages}{3--17}.
\newblock


\bibitem[Nguyen et~al\mbox{.}(2016)]%
        {nguyen2016msmarco}
\bibfield{author}{\bibinfo{person}{Tri Nguyen}, \bibinfo{person}{Mir
  Rosenberg}, \bibinfo{person}{Xia Song}, \bibinfo{person}{Jianfeng Gao},
  \bibinfo{person}{Saurabh Tiwary}, \bibinfo{person}{Rangan Majumder}, {and}
  \bibinfo{person}{Li Deng}.} \bibinfo{year}{2016}\natexlab{}.
\newblock \showarticletitle{MS MARCO: A Human Generated MAchine Reading
  COmprehension Dataset}.
\newblock  (\bibinfo{date}{November} \bibinfo{year}{2016}).
\newblock


\bibitem[Reimers and Gurevych(2019)]%
        {reimers-2019-sentence-bert}
\bibfield{author}{\bibinfo{person}{Nils Reimers} {and} \bibinfo{person}{Iryna
  Gurevych}.} \bibinfo{year}{2019}\natexlab{}.
\newblock \showarticletitle{Sentence-BERT: Sentence Embeddings using Siamese
  BERT-Networks}. In \bibinfo{booktitle}{\emph{Proceedings of the 2019
  Conference on Empirical Methods in Natural Language Processing}}.
  \bibinfo{publisher}{Association for Computational Linguistics}.
\newblock


\bibitem[Robertson et~al\mbox{.}(1994)]%
        {bm25original}
\bibfield{author}{\bibinfo{person}{Stephen~E. Robertson},
  \bibinfo{person}{Steve Walker}, \bibinfo{person}{Susan Jones},
  \bibinfo{person}{Micheline Hancock-Beaulieu}, {and} \bibinfo{person}{Mike
  Gatford}.} \bibinfo{year}{1994}\natexlab{}.
\newblock \showarticletitle{Okapi at TREC-3.}. In
  \bibinfo{booktitle}{\emph{TREC}} \emph{(\bibinfo{series}{NIST Special
  Publication}, Vol.~\bibinfo{volume}{500-225})},
  \bibfield{editor}{\bibinfo{person}{Donna~K. Harman}} (Ed.).
  \bibinfo{publisher}{National Institute of Standards and Technology (NIST)},
  \bibinfo{pages}{109--126}.
\newblock


\bibitem[Santhanam et~al\mbox{.}(2022)]%
        {santhanam2022colbertv2}
\bibfield{author}{\bibinfo{person}{Keshav Santhanam}, \bibinfo{person}{Omar
  Khattab}, \bibinfo{person}{Jon Saad-Falcon}, \bibinfo{person}{Christopher
  Potts}, {and} \bibinfo{person}{Matei Zaharia}.}
  \bibinfo{year}{2022}\natexlab{}.
\newblock \showarticletitle{ColBERTv2: Effective and Efficient Retrieval via
  Lightweight Late Interaction}. In \bibinfo{booktitle}{\emph{Proceedings of
  the 2022 Conference of the North American Chapter of the Association for
  Computational Linguistics: Human Language Technologies}}.
  \bibinfo{pages}{3715--3734}.
\newblock


\bibitem[Thakur et~al\mbox{.}(2021)]%
        {thakur2021beir}
\bibfield{author}{\bibinfo{person}{Nandan Thakur}, \bibinfo{person}{Nils
  Reimers}, \bibinfo{person}{Andreas R{\"u}ckl{\'e}}, \bibinfo{person}{Abhishek
  Srivastava}, {and} \bibinfo{person}{Iryna Gurevych}.}
  \bibinfo{year}{2021}\natexlab{}.
\newblock \showarticletitle{{BEIR}: A Heterogeneous Benchmark for Zero-shot
  Evaluation of Information Retrieval Models}. In
  \bibinfo{booktitle}{\emph{35th Conference on Neural Information Processing
  Systems Datasets and Benchmarks Track (Round 2)}}.
\newblock


\bibitem[Wang et~al\mbox{.}(2020)]%
        {wang2020minilm}
\bibfield{author}{\bibinfo{person}{Wenhui Wang}, \bibinfo{person}{Furu Wei},
  \bibinfo{person}{Li Dong}, \bibinfo{person}{Hangbo Bao}, \bibinfo{person}{Nan
  Yang}, {and} \bibinfo{person}{Ming Zhou}.} \bibinfo{year}{2020}\natexlab{}.
\newblock \showarticletitle{Minilm: Deep self-attention distillation for
  task-agnostic compression of pre-trained transformers}.
\newblock \bibinfo{journal}{\emph{Advances in Neural Information Processing
  Systems}}  \bibinfo{volume}{33} (\bibinfo{year}{2020}),
  \bibinfo{pages}{5776--5788}.
\newblock


\bibitem[Xiong et~al\mbox{.}(2021)]%
        {xiong2021approximate}
\bibfield{author}{\bibinfo{person}{Lee Xiong}, \bibinfo{person}{Chenyan Xiong},
  \bibinfo{person}{Ye Li}, \bibinfo{person}{Kwok-Fung Tang},
  \bibinfo{person}{Jialin Liu}, \bibinfo{person}{Paul Bennett},
  \bibinfo{person}{Junaid Ahmed}, {and} \bibinfo{person}{Arnold Overwijk}.}
  \bibinfo{year}{2021}\natexlab{}.
\newblock \showarticletitle{Approximate Nearest Neighbor Negative Contrastive
  Learning for Dense Text Retrieval}. In
  \bibinfo{booktitle}{\emph{International Conference on Learning
  Representations}}.
\newblock


\bibitem[Zhao et~al\mbox{.}(2021)]%
        {zhao2021sparta}
\bibfield{author}{\bibinfo{person}{Tiancheng Zhao}, \bibinfo{person}{Xiaopeng
  Lu}, {and} \bibinfo{person}{Kyusong Lee}.} \bibinfo{year}{2021}\natexlab{}.
\newblock \showarticletitle{SPARTA: Efficient Open-Domain Question Answering
  via Sparse Transformer Matching Retrieval}. In
  \bibinfo{booktitle}{\emph{Proceedings of the 2021 Conference of the North
  American Chapter of the Association for Computational Linguistics: Human
  Language Technologies}}. \bibinfo{pages}{565--575}.
\newblock


\end{thebibliography}


\end{document}